\documentclass{iaus}
\usepackage{graphicx}
\usepackage{natbib}
\usepackage{epsfig}

\title[IAUS 289.~~The 2MASS Tully-Fisher Survey] 
{The 2MASS Tully--Fisher Survey : \\Mapping the Mass in the Universe}

\author[T.~Hong et al.]   
{T.~Hong,$^{1,2,3}$ L.~Staveley-Smith,$^{2,3}$ K.~Masters,$^4$ C.~Springob,$^{5,2,3}$ 
L.~Macri,$^6$ B.~Koribalski,$^7$ H.~Jones,$^8$ \and T.~Jarrett$^9$}

\affiliation{$^1$National Astronomical Observatories, Chinese Academy
  of Sciences, 20A Datun Road, Chaoyang District, Beijing 100012,
  China.\\email: {\tt bartonhongtao@gmail.com}\\ 
$^2$International
  Centre for Radio Astronomy Research, M468, University of Western
  Australia, Crawley, WA 6009, Australia\\ 
$^3$ARC Centre of
  Excellence for All-sky Astrophysics (CAASTRO)\\ 
$^4$Institute for
  Cosmology and Gravitation, University of Portsmouth, Dennis Sciama
  Building, Burnaby Road, Portsmouth PO1 3FX\\ 
$^5$Australian Astronomical Observatory, PO Box 296, Epping, NSW 1710, Australia\\ 
$^6$George P. and Cynthia Woods Mitchell Institute for Fundamental Physics and 
Astronomy, Department of Physics and Astronomy, Texas A\&M University, 4242 TAMU, 
College Station, TX 77843, USA\\ 
$^7$CSIRO Astronomy \& Space Science, PO Box 76, Epping, NSW 1710, Australia\\ 
$^8$School of Physics, Monash University, Clayton, VIC 3800, Australia\\ 
$^9$Astronomy Department, University of Cape Town, Private Bag X3. 
Rondebosch 7701, South Africa\\ }

\pubyear{2012}
\volume{289}  
\pagerange{}
\jname{Advancing the physics of cosmic distances}
\editors{Richard de Grijs \& Giuseppe Bono, eds.}
\begin{document}

\maketitle

\begin{abstract}
The 2{\sc mass} Tully-Fisher Survey (2MTF) aims to measure 
Tully-Fisher (TF) distances for all bright inclined spirals 
in the 2{\sc mass} Redshift Survey (2MRS) using high quality 
HI widths and 2{\sc mass} photometry. Compared with previous 
peculiar velocity surveys, the 2MTF survey provides more 
accurate width measurements and more uniform sky coverage, 
combining observations with the Green Bank, Arecibo and Parkes 
telescopes. With this new redshift-independent distance 
database, we will significantly improve our understanding 
of the mass distribution in the local universe. 
\end{abstract}

\firstsection 
\section{Introduction}
The Tully-Fisher relation is an empirical relation between the luminosity
and rotational velocity of spiral galaxies \citep{tf1977}. As a
secondary distance indicator, the Tully-Fisher relation is a good tool for
measuring the redshift-independent distances of local spiral
galaxies. With these redshift independent distances, we can calculate
the peculiar velocity field, which in turn allows us to trace the mass
distribution of the local universe.

In the last few decades, a number of Tully-Fisher surveys have been conducted, 
including those described in 
\citet{ghh+1997,smh+2007,tsk+2008}, which are typically limited 
by source selection criteria and sky-coverage. For instance, the SFI++ 
survey \citep{smh+2007}, which is the largest Tully-Fisher survey to date, was 
selected optically in I-band and can only cover Galactic latitudes down to $|b|=15^{\circ}$.  This 
sky-coverage is a significant limitation on the measurement of an all-sky peculiar 
velocity field, especially in the Zone of Avoidance (ZoA; the part of the 
sky which is difficult to be observed because of dust and source crowding 
in our own Galaxy). Thus, a new 
all-sky Tully-Fisher survey with uniform source selection and sky-coverage 
is needed for the study of the local peculiar velocity field.  This new survey 
will improve our model of the mass distribution of the local Universe.

\section{2{\sc mass} Tully-Fisher Survey}

The 2{\sc mass} Tully-Fisher Survey (2MTF), which is based on a source list selected from the 2
Micron All-Sky Survey (2{\sc mass}), will provide better statistics and more
even sky coverage than previous surveys, in particular greatly reducing the impact of the
ZoA. This survey will make use of existing high quality
rotation widths, new HI widths, and 2{\sc mass} photometry to measure
Tully-Fisher distances for all bright inclined spirals in the
2{\sc mass} Redshift Survey \citep[2MRS,][]{hmm+2012}.

\subsection{Source selection}
\begin{figure*}
\centering
\includegraphics[width=0.40\columnwidth, angle=-90]{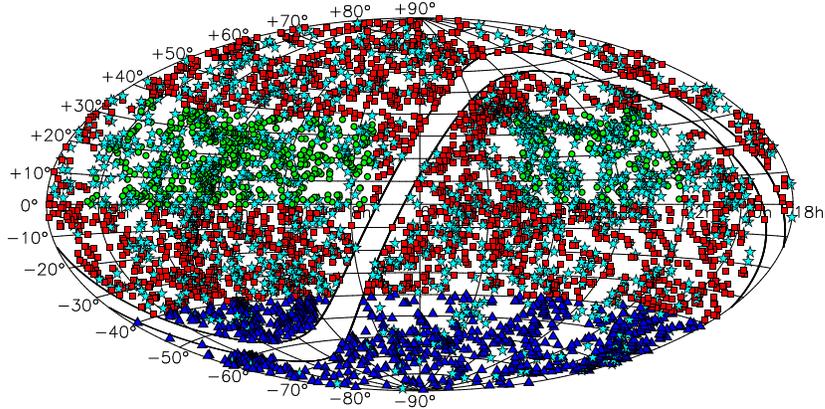}
\caption{The distribution of 6,000 2MTF target galaxies. The red
  squares indicate the GBT observed galaxies, Parkes observed galaxies
  are plotted by blue triangles, the green circles show the ALFALFA
  galaxies, and the cyan stars are galaxies with archival data. The thick lines 
  trace the galactic latitudes $b=5^{\circ}$ and $b=-5^{\circ}$.}
\label{fig:sky_dis}
\end{figure*}

To minimize the error in the final Tully-Fisher distances, we selected
only bright and highly-inclined galaxies from 2MRS with a limit of
$K_{s}=11.25$ mag, $cz<10,000$ km s$^{-1}$, and axial ratio $b/a <
0.5$. The target sample contains approximately 6,000 galaxies, which covers 
more than 90\% of the whole sky, where the missing 10\% is due to the Milky Way. 
35\% of these galaxies have rotation width measurements
for Tully-Fisher distances already available from the literature, but with very uneven sky coverage. Figure~\ref{fig:sky_dis} shows the distribution of 6,000 target galaxies.

\subsection{Current Data Status}
Our new HI line observations contain about 1,300 galaxies observed with high
velocity resolution, conducted with the Green Bank Telescope (GBT) and the Parkes
Telescope between Feb 2006 to Feb 2012. When complete in 2012, the
ongoing Arecibo Legacy Fast ALFA  
survey \citep[ALFALFA,][]{ghk+2005} survey will also provide high velocity resolution
widths for all HI rich galaxies in the high galactic latitude Arecibo
sky.

\subsubsection{New Observations in GBT and Parkes Telescope}

In the northern sky ($\delta > - 40^{\circ}$), we observed about 1,000
galaxies using the GBT in the 06A, 06B and 06C semesters. Observations
were done in position switched mode, always in pairs of ~5 mins ON/OFF
with a 12.5 MHz bandwidth and 8192 channels.

In the sky south of $- 40^{\circ}$, 305 galaxies which meet our
selection criteria were considered to be observable at Parkes without confusion.  In six
semesters (06OCTS, 07APRS, 07OCTS, 08APRS, 08OCTS and 11OCTS) at
Parkes, we have observed all of these 305 galaxies using the 20cm
multi-beam receiver in beam switching mode with a bandwidth of 8 MHz and
1024 spectral channels.

To obtain the lowest possible error in Tully-Fisher distances, we have to minimize
the error (less than 10\%) on the measurements of rotation widths. We
require S/N $>$ 10 to measure the widths to better than 10\%. From
these new observed HI lines, we got 386 high quality HI width
measurements from GBT data, and also obtained 152 high quality HI spectra
from Parkes.
\begin{figure*}
\centering
\includegraphics[width=0.40\columnwidth, angle=-90]{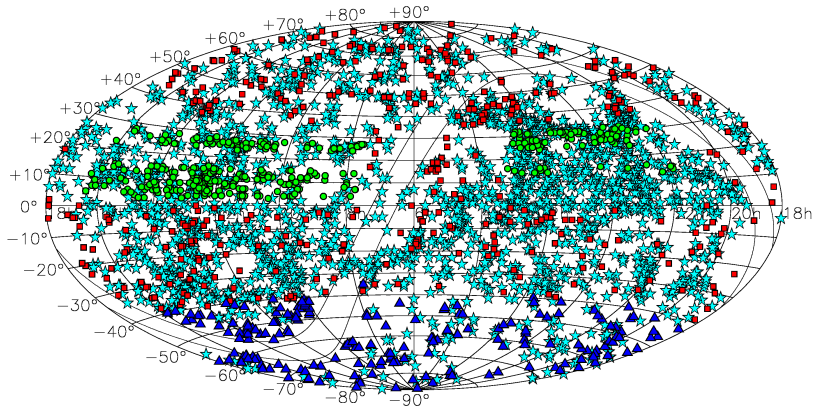}
\caption{The distribution of the final sample of 3,000 2MTF galaxies 
(pending the addition of the remaining ALFALFA galaxies). All symbols share 
the same meaning as in Figure~\ref{fig:sky_dis}}
\label{fig:sky_dis_useful}
\end{figure*}
\subsubsection{ALFALFA 40\% Data}
ALFALFA is a large blind HI
survey being undertaken with the Arecibo telescope. It covers the high galactic latitude Arecibo
sky, and will be completed in late 2012. More than 30,000 extragalactic HI
sources will be detected by ALFALFA with redshift up to $z \sim 0.06$.

40\% of the ALFALFA data ($\sim 15,900$ HI sources) have been published by
\citet{hgm+2011}.  After cross-matching with our 2MTF target sample, we
found 357 useful widths for our Tully-Fisher calculations. We
still await the full data release, so that we may complete our sample.

\subsubsection{Archived Data}
The archived HI widths are mainly from the SFI++ database
\citep{shgk2005}. Besides the SFI++ data, we also selected HI widths from more than 10
additional catalogs in the literature. About 2,000 archived HI data sources matched our 2MTF
target sample, and 1,800 widths have accuracy better than 10\%.\\

\begin{figure*}
\centering
\includegraphics[width=0.30\columnwidth, angle=-90]{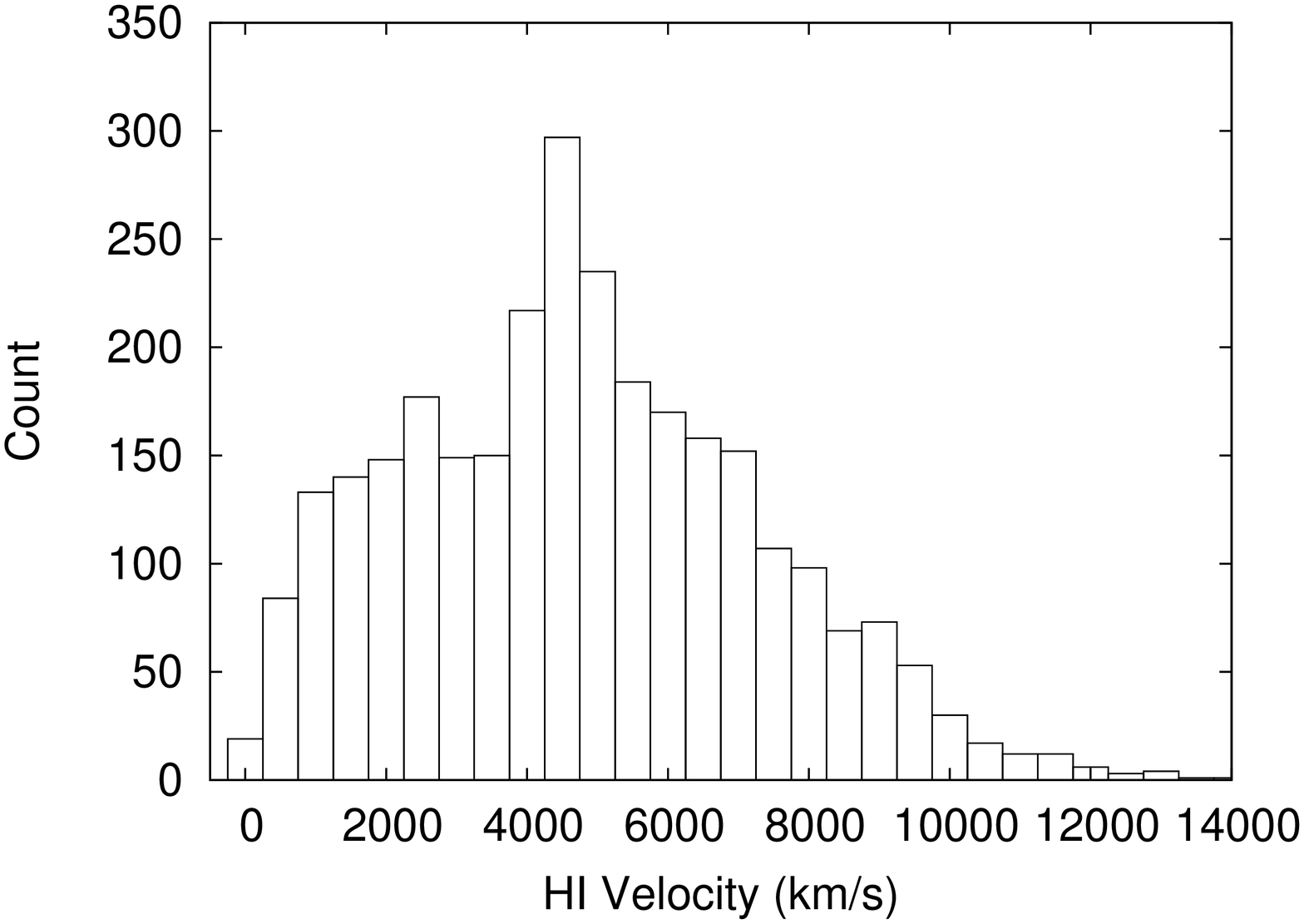}
\includegraphics[width=0.30\columnwidth, angle=-90]{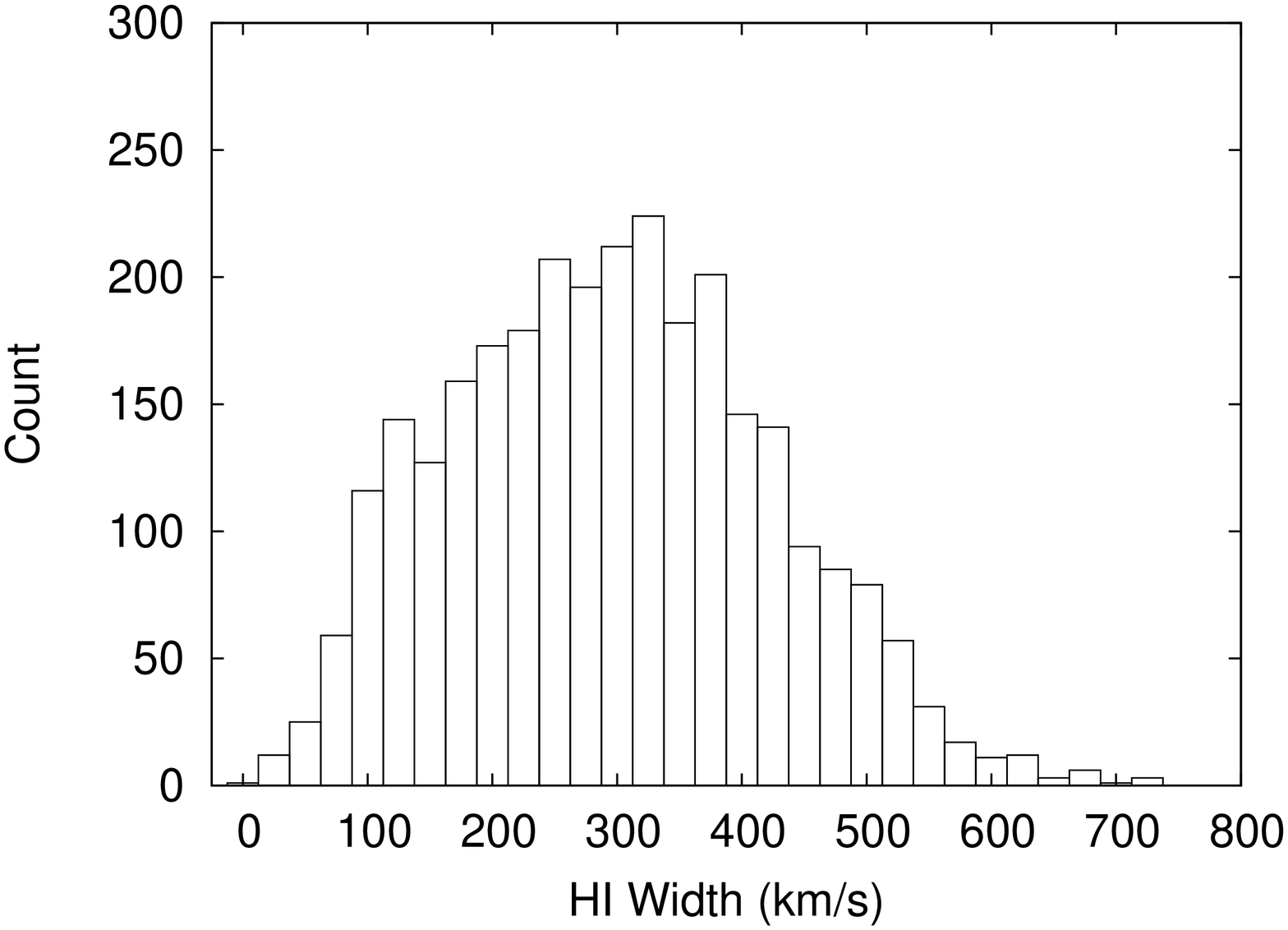}
\caption{The central velocity and HI width histograms of the final 2MTF sample.}
\label{fig:his_vel_wid}
\end{figure*}

Our final sample (pending the addition of the remaining ALFALFA galaxies) 
includes roughly 3,000 useful HI widths with uniform distribution
in total.  This includes all of the newly observed galaxies, the ALFALFA 40\% data and
the archival data. The spectroscopic data will be published and made available online shortly. 
The data releasing and analyzing papers are also in prepare 
(Masters et al. 2012, to be submitted; Hong et al. 2012, to be submitted).
Figure~\ref{fig:sky_dis_useful} shows the sky distribution 
of the 2MTF final sample, and Figure~\ref{fig:his_vel_wid} shows the 
histograms of the galaxies' central velocities and HI widths.

\subsection{New Near-IR Tully-Fisher Template}
A universal Tully-Fisher calibration in the near-infrared bands is very
important for the 2MTF project. \citet{msh2008} built a new
Tully-Fisher template in the 2{\sc mass} K, H and J bands using the ``basket of
clusters" method \citep{ghh+b1997}, with a sample containing 888
galaxies in 33 clusters.

\citet{msh2008} also split the full sample into three sub-samples by
galaxy morphology, and determined that the Tully-Fisher relation
depends on galaxy morphology in all three 2{\sc mass} bands, with later type
galaxies having a steeper Tully-Fisher slope and a fainter zero point than earlier type galaxies. \cite[see][figure
4]{msh2008}.

\citet{msh2008} corrected the final relation to that for Sc galaxies,
and this relation will be used as the universal template for the 2MTF
calculations.

\section{Conclusion}
2MTF project is an all-sky Tully-Fisher survey, will measure
Tully-Fisher distances of all bright highly-inclined galaxies in the local
universe. Comparing with previous Tully-Fisher surveys, the 2MTF
project provides more even sky-coverage and a smaller ``Zone of
Avoidance", and will be a better sample for measuring the peculiar
velocity field in the local universe.  The final 2MTF sample contains about
3,000 high quality HI widths, all selected from the 2{\sc mass} Redshift Survey. The sample contains three parts,
the new observed HI widths by our group using the GBT and Parkes
telescope, the ALFALFA HI widths and the archived high quality 
HI widths from the literature. A new calibration of the
Near-IR Tully-Fisher relation has been made and published. The 2MTF
project will help us to study and understand the peculiar velocity
field in the local universe, using the 2MTF data, we will provide better constraints 
on the local bulk flow, dipole motion and mass distribution.

\bibliographystyle{apj}
\bibliography{bibfile}

\end{document}